# THE IMPACT OF TEACHING TWO COURSES (ELECTRONIC CURRICULUM DESIGN AND MULTIMEDIA) ON THE ACQUISITION OF ELECTRONIC CONTENT DESIGN SKILLS

Natheer K Gharaibeh *, Mohareb A Alsmadi **

Albalqa Applied University, Jordan.
*Natheer_garaybih@yahoo.com, Mohareb_smadi@yahoo.com **

## ABSTRACT

*The use of Multimedia applications in Learning provides useful concepts for Instructional Content Design. This study aimed to investigate the effect of design electronic curriculum and multimedia applications on acquiring e-content design skills, and improving their attitudes towards e-learning. To achieve the objective of the study, the researchers developed a test to measure the efficiencies of designing electronic content and the measure of attitudes towards e-learning, The results showed that study of both courses contributed positively to the acquisition of design skills of e-content , The results revealed that there are statistical significant differences between the scores of the students in the two applications (pre and post) on the total score of the attitude measure and three areas of it.*

## 1. INTRODUCTION

With the emergence of globalization and the rapid growth of electronic information society, Accompanied with the emergence of the concepts of e-schools and virtual classes, the educational systems in Arab countries become in the face of many of the enormous challenges that must be addressed in educational thinking so new strategies can be advanced in order to prepare the future generations to have skills in dealing with the variables of the current century.

The e-curricula and e-learning is considered the most important of the basic fundamentals in Jordan and its institutions in building their future in the information age and Electronics.The idea of electronic curricula has emerged to the extent that some experts predict that the electronic school will be the most ideal and popular method for education and training in the near future

E-learning is a broad term that covers a wide range of educational materials that can be provided on CD-ROM or through a local area network (LAN) or the Internet. It includes computer-based training, and web based training, electronic performance support systems, distance learning, online learning, e-tutoring [1]. In this context we understand the term 'multimedia' as software that combines a range of media Sound, image and movement, text, drawing and high quality video all working under the control of the computer at one time [2].

The e-learning is one of the most modern methods of learning, it increases the effectiveness of learning to a great extent and reduces the time required for training and reduces the cost of training [3] , e-learning supports online work interviews and lively discussions on the network, provides a modern information consistent with the needs of learners, and provides simulation programs , motion pictures, events and interactive exercises and practical applications [4] .

Among the benefits of e-learning also the ability to meet the needs of individual learners so that individuals learn by their selves, saves the cost of training (accommodation, travel, books) , Better retain of the information and access to information in a timely manner, the Speed of updating information in the network , Standardization of content and information to all users , Improve cooperation and interaction between students, and reduces the feeling of the student embarrassed in front of his colleagues when he committed the error (Codone, 2001).

Hence, the role of the teacher has changed from the past, The evolution of information and communication technology added new burdens to many teachers today, Which has become imperative for him to deal with modern technology and employ multimedia in the teaching process to help students achieve educational outcomes [2]. Owning the teacher the design skills of e-content and multimedia production and use will improve student achievement of knowledge and expertise, and keep them out, and makes them more able to use this knowledge and its application in the attitudes of working life .

Despite the development of preparation methods for teachers in colleges of education in Jordanian universities , the interest in employing multimedia in university teaching is still modest. The process of preparation and development of programs and platforms for e-learning is considered the most important requirements of the application of e-learning, where this process requires a great effort, experts and specialists in the design and programming.

## 2. BACKGROUND OF THE STUDY

### 2.1 Problem of the Study

Providing students with the expertise of scientific and practical skills in the design of electronic content , and employing multimedia requires Good preparation and practical competencies , the formation of positive attitudes towards the employment of multimedia in the design of teaching [5] . The use of technology in educational process is no longer a manifestation of academic luxury, but is necessary and inevitable concern required by the times and circumstances [6] , Jibrin Has pointed out that the most important factors that reduce the use of multimedia is the lack of teacher training in Jordan.

Through teaching of the for the Master's program in information and communication technology in education, and through field visits to the students who are teachers in schools , the researchers noticed that there is a reluctance to employ multimedia in the learning process; this situation can be result of one or more of these reasons:

1- Lack of awareness of its importance
2- Lack of conviction
3- Not possessing the competences of their design and implementation

Since the students are one of the input elements of education system, it is necessary to find out : how to acquire e-content design skills, trends and areas of teaching , `constraints applied, and how they have the skills to employ multimedia in the learning process.

For that reason the researchers wanted to check it all, by identifying the impact of teaching electronic curriculum design and multimedia applications on the acquisition of electronic content design skills Specifically, Two questions emerge and are central to this investigation:

1. To what extent teaching the two courses (electronic curriculum design, multimedia applications in teaching) can contribute to making students acquire the skills of electronic content design.
2. Does the students' attitudes for the design of electronic content after studying two courses (electronic curriculum design, multimedia applications in teaching)  different from their attitudes prior to study.

## 2.2 Objectives of the study:

The study aimed to investigate the effect of teaching the two courses (electronic curriculum design, multimedia applications in teaching)  of students in the skills of e-content design and their desire to learn and apply it in the field. And investigate the change in students' attitudes towards it.

## 2.3 The limits of the study:

The study sample was limited to graduate students in information and communication technology in education at the University of Al al-Bayt (Part of Ajloun) in the second semester of 2012, as the results of the study is determined by the nature of the tools used, and significance of sincerity and persistence.

## 2.4 Procedural definitions:

**Digital curriculum :**

Is an approach based on the integration of content, pedagogy and environment designed for, delivered by and supported with digital means within a digital frame of thinking. student studies its contents technology and interactive with the faculty member at any time and any place wants, the content includes electronic Hyper text and Hyper Links  and Hyper Media and the design of graphics animation and simulation  of skills and examples.

**Two courses (electronic curriculum design, multimedia applications in teaching)**

Which Is represented in this study as the plan adopted at the University of Al al-Bayt in the level of master of information and communication technology in education, containing specific goals in advance and four units for each course, and a group educational activities, and activities

The International Journal of Multimedia & Its Applications (IJMA) Vol.4, No.6, December 2012

calendar, designed to provide students with information and expertise of some multimedia applications, and produce and use , and the skills, digital curriculum design models.

## 2.5 Sample of the study

The study population consisted of all graduate students in information and communication technology in education at the University of Al al-Bayt and cooperation with the Arab Group for Education and Training, the number (84) divided into three parts: the Part of Al al-Bayt, Part of Karak and Part of Ajloun, and the study sample is formed of Students from Part of Ajloun ,15 students were chosen in deliberate way because the researchers are the instructors of this part, and formed a rate (18%) of the total study population.

## 3. METHODOLOGY

The study followed the descriptive and analytical Methodology

## 3.1 Instruments of the study:

The researchers used two tools to achieve the objective of the study :

1- Achievement test
To measure the skills of designing electronic content and employ multimedia in it, it was constructed of 50 paragraphs of type multiple choice, the coefficient of difficulty and coefficient of discrimination of each paragraph have varied degrees of difficulty between (0.54-0.82), and varied degrees of discrimination between (0.15 -0.37) , and these grades are considered acceptable for the purposes of the study.

2 - Attitudes Questionnaire
Design of electronic content was developed after it has been reviewed through several measures of trends in the field of electronic teaching and multimedia production and development, and the obstacles they face, both locally and globally [7] [8] [9] , the Questionnaire constructed from 64 questions according to the Likert scale . The signs of the scale ranged between per capita (64-320), and is considered the mark inclined 192 above the equivalent of 60% of the total marks, the mark interval between the linear trend is positive and the negative of the responder.

## 3.2 Validity test

We build the tools of the study after analyzing the content of the two courses and the formulation of educational goals and review the tools of previous studies, and it has been given – with the objectives of the two courses and content and their plans - to 10 arbitrators specialists in the field of curriculum and technology education in Balqa Applied University, University of Al al-Bayt, we amended some of the paragraphs based on their suggestions, this was seen as an indication of the validity of the study tools.

### 3.3 Reliability study tools:

coefficient of internal consistency of the test was calculated using equation (Kuder-Richardson: Kr-20) and amounted to (0.84) , this value is considered relatively high and reliable to access the stabilized data.

Have also been identified stability measure by calculating the stability of the total degree of the scale , the correlation coefficient calculated using Cronbach alpha equation to measure the tribal was (0.89), and (0.88) to measure the post, and the standard became applicable in its final form. To verify the Reliability of the analysis, the two researchers make individual analysis and evaluation of student projects, and then find the proportion of agreement between analysts and reached 88%.

## 4. THE RESULTS OF THE STUDY

To answer the first question: the means of the marks was calculated for pre test and post test, to determine the degree of possessing cognitive competencies for students designing electronic content before and after the study the two courses. And table (1) shows the results

Table(1) Means, standard deviations and the results of T test of the interrelated samples (Correlated Groups) to measure the difference between the arithmetic mean of the pre test and post test

| Type of test | N | Mean | Std. Deviation | Std. Error Mean | T | df. | Sig. (2-tailed) |
|---|---|---|---|---|---|---|---|
| Pre- test | 15 | 22.53 | 7.97197 | 2.05836 | 5.411 | 14 | .000 |
| Post test | 15 | 34.20* | 2.48424 | 0.64143 | | 14 | |

* the mean form 50

It was shown in Table (1) that the arithmetic mean of the post test(34.20) is greater than the arithmetic mean of the pre test(22.53). Also, The value of (t) calculated (5.411) and the level of significance (0.000), which reveals a significant difference ($\alpha = 0.01$) between the pretest and post test in favor of the post test, which indicates that students master the cognitive skills of 68.4% in the design of electronic content, and that the study of the two courses contributed positively to the acquisition of electronic content design skills.

To answer the second question: "Does the students' attitudes for the design of electronic content after studying two courses (electronic curriculum design, multimedia applications in teaching) different from their attitudes prior to study", the researchers use the means of the marks and mathematical test (t) for (Correlated Groups) to measure the difference between the arithmetic mean of estimates members of the study sample on the scale as a whole and the four attitudes for teaching before and after, as shown in Table (2):

Table(2) the means of the marks and mathematical test (t) for (Correlated Groups)

|  | Type-test | N | Mean | Std. Deviation | Std. Error Mean | T | Sig. (2-tailed) |
|---|---|---|---|---|---|---|---|
| Total | Pretest | 15 | 201.0667 | 13.01903 | 3.36150 | -6.971 | .000 |
|  | Posttest | 15 | 238.9333 | 16.52473 | 4.26667 |  |  |
| domain | Pretest | 15 | 51.8667 | 4.89704 | 1.26441 | -8.313 | .000 |
|  | Posttest | 15 | 77.0000 | 10.63686 | 2.74643 |  |  |
| use | Pretest | 15 | 69.0000 | 5.68205 | 1.46710 | -3.141 | .004 |
|  | Posttest | 15 | 77.5333 | 8.85492 | 2.28633 |  |  |
| Difficulties | Pretest | 15 | 72.2667 | 8.97191 | 2.31654 | 1.470 | .153 |
|  | Posttest | 15 | 67.9333 | 7.06568 | 1.82435 |  |  |
| Skills | Pretest | 15 | 7.9333 | 3.10453 | .80159 | -8.038 | .000 |
|  | Posttest | 15 | 16.4667 | 2.69568 | .69602 |  |  |

It was shown in Table (2) that the attitudes of the participants comes at first; the responding means after studying the two courses is (77.91) while the means before studying the two courses is (68.77) followed by the attitudes towards the fields of e-learning use of (77.89, 51.89)after and before studying the two courses respectively. the difficulty in using e-learning after studying the two courses(68.09) is less than before studying the two courses (72.20). Finally, the means score (16.42) of their employment of multimedia in teaching after studying the two courses is higher than the means score(8.00) before studying the two courses.

It was shown in Table (2) the existence of statistical significant differences between scores by students in the pre and post applications, on the total score of the scale and the three areas of it, namely: the trend towards the areas of e-learning, the trend towards the use of e-learning, skills and employment of multimedia. The table not revealed existence of a statistical significant difference between students scores on the third area: the trend towards difficulties the use of e-learning between the application pre and post ; it could be argued that students attitudes towards the design of electronic content and employ multimedia which increased essentially, however it was in the pre test tend towards the positive, but it is becoming more and more positive in the post test .

## 5. DISCUSSION

The results of the study indicated that the study of the two courses (curriculum design, electronic multimedia applications in teaching) contributed in giving students high capabilities of design efficiencies of electronic content. The reason for this is due to the nature of design, design models of electronic content., the procedural steps of the techniques employed in the process of teaching and learning in a way Trotted out to students, simplifies the concepts, and helps to understand the article and remember, Students do hands-on practice for the design of electronic unit depending on the design of e-content models and employ multimedia interaction design, this has contributed to the development of theoretical and practical experience for students , and increase the desire to take more of their knowledge and sciences, including containing models and applications of multimedia in teaching offers teaching positions in the live image with the integrated voice and movement, This raises the attention of learners and take them out for the traditional context. All this contributed to acquire a high proportion of cognitive skills, and provided them with information and the necessary skills and enshrined in their minds through their study of the two courses.

The reason for the lack of students knowledge for all multimedia skills is the lack of some materials and software required for multimedia production, This result agrees with the results of (Zagal, 1991) which showed that a technology education course which was applied to the graduate students has impact in the development of scientific skills and cognitive skills for students.

The study results indicated that the study of the two courses have contributed to increase the positive trends for the students to design e-content and employ multimedia in teaching, especially on the three trends:

1-the trend towards the applications of electronic teaching,
2-the trend towards the use of electronic teaching,
3-the trend towards employment of multimedia in electronic teaching.

The researchers noticed that students' attitudes before the study the two courses tend to be positive, the reason for this was their awareness of the importance, attract their attention and consolidate the ideas and concepts in their minds, this make the attitudes positive towards electronic teaching and employing multimedia in it.

It seems that the study of the two courses (curriculum design, electronic multimedia applications in teaching) has increased the students' attitudes towards the design of electronic content and employ multimedia in which in general, specifically the three trends the applications , the use and employment of electronic teaching, this result has attributed to the main role of multimedia, which work on providing electronic content in an interactive and simple, enabled the students to identify the applications of multimedia they did not practiced before, furthermore the practical training in the production of projects for the design of electronic units and employ multimedia in classroom, made them feel that they walk with technological development witnessed by the world, the students' considered this as educational renewal and an update to Many curricula and methods of teaching.